\documentclass[11pt,dvips]{article}
\usepackage{eurosym}
\usepackage{graphicx}
\usepackage{amsmath}
\usepackage{amssymb}
\usepackage{latexsym}
\usepackage{color}
\usepackage{epstopdf}

\input{tcilatex}
\begin{document}

\thispagestyle{empty}

\begin{center}
\vspace{1.8cm}

%%%%%%%%%%%%%%%%%%%%%%%%%%%%%%%%%%%%%%%%%%%%%%%%%%%%%%%%%%%%%%%%%%%%%%%%%%%%%%%%%%%%%%%%%%%%%%%%%%%%%%%%%%%%%%%%%%%%%%%
{\Large \textbf{Dynamical Gaussian quantum steering in optomechanics}}
%%%%%%%%%%%%%%%%%%%%%%%%%%%%%%%%%%%%%%%%%%%%%%%%%%%%%%%%%%%%%%%%%%%%%%%%%%%%%%%%%%%%%%%%%%%%%%%%%%%%%%%%%%%%%%%%%%%%%

\vspace{1.5cm}

\textbf{Jamal El Qars}$^{a}${\footnote{%
email: \textsf{j.elqars@gmail.com}}}, \textbf{Mohammed Daoud}$^{b,c,d}${\footnote{%
email: \textsf{m$_{-}$daoud@hotmail.com}}} and \textbf{Rachid Ahl Laamara}$%
^{a,e} ${\footnote{%
email: \textsf{ahllaamara@gmail.com}}}

\vspace{0.5cm}

$^{a}$\textit{LPHE-MS, Faculty of Sciences, Mohammed V University of Rabat,
Rabat, Morocco}\\[0.5em]
$^{b}$\textit{Max Planck Institute for the Physics of Complex Systems,
Dresden, Germany}\\[0.5em]
$^{c}$\textit{Abdus Salam International Centre for Theoretical Physics,
Miramare, Trieste, Italy}\\[0.5em]
$^{d}$\textit{Department of Physics, Faculty of Sciences, University Hassan
II, Casablanca, Morocco}\\[0.5em]
$^{e}$\textit{Centre of Physics and Mathematics (CPM), Mohammed V University
of Rabat, Rabat, Morocco}\\[0.5em]

\vspace{3cm} \textbf{Abstract}
\end{center}

\baselineskip=18pt Einstein-Podolski-Rosen steering is a form of quantum
correlation exhibiting an intrinsic asymmetry between two entangled systems.
In this paper, we propose a scheme for examining dynamical Gaussian quantum
steering of two mixed mechanical modes. For this, we use two spatially
separated optomechanical cavities fed by squeezed light. We work in the
resolved sideband regime. Limiting to the adiabatic regime, we show that it
is possible to generate dynamical Gaussian steering via a quantum
fluctuations transfer from squeezed light to the mechanical modes. By an
appropriate choice of the environmental parameters, one-way steering can be
observed in different scenarios. Finally, comparing with entanglement -
quantified by the Gaussian R\'{e}nyi-2 entropy -, we show that Gaussian
steering is strongly sensitive to the thermal effects and always upper
bounded by entanglement degree.
%%%%%%%%%%%%%%%%%%%%%%%%%%%%%%%%%%%%%%%%%%%%%%%%
\newpage

\section{Introduction}

Einstein-Podolsky-Rosen or EPR steering \cite{EPR} is one of several aspects
of \textit{inseparable} quantum correlations such as entanglement \cite%
{Erwin and Horodecki} and Bell's non-locality \cite{Bell and
Clauser,Wiseman,Wollmann,ANGB}. In the hierarchy, quantum steering sits
between entanglement and Bell's non-locality, as the asymmetric roles (not
exchangeable) played by two entangled observers Alice and Bob makes it
distinct. This phenomenon, which is the heart of the EPR paradox \cite%
{Reid(1)}, was firstly introduced by Schr\"{o}dinger \cite{Schrodinger(2)}
to reveal the non-locality in the EPR states and to highlight that such
classes of quantum states are implicitly entangled. In quantum information
theory, the distinctive feature of quantum steering compared to the other
phenomena is its directionality \cite{Wiseman}. Indeed, for two observers,
Alice and Bob, who jointly share an entangled state, steerability allows
Alice (for instance) by performing local measurement to non-locally affect
(i.e., steer) Bob's states \cite{Feng Li}. In other words, quantum steering
corresponds to an entanglement verification task in which one party is
untrusted \cite{Wiseman}. In fact, if Alice can steer Bob's states, then she
is able to convince Bob (who does not trust Alice) that their shared state
is entangled by performing local measurements and classical communication
(LMCC) \cite{Wiseman,Kogias}.\newline
Reid later proposed experimental criteria for detection of the EPR paradox
for continuous-variable systems (CVs) \cite{Reid(2)}; where the first
experimental observation of this effect has been achieved by Ou \textit{et al%
} \cite{Ou}, and was followed by a great number of recent works \cite%
{Reid(1),G(2)}. On the other hand, it has been shown by Wiseman \textit{et al%
} \cite{Wiseman}, that under Gaussian measurements, violation of the Reid
criteria is a genuine demonstration of EPR steering. Interestingly enough,
Wiseman \textit{et al} \cite{Wiseman} have been already raised an important
question of whether there exist entangled states which are one-way
steerable, i.e., Alice can steer Bob's state but it is impossible for Bob to
steer the state of Alice even though they are entangled. Thanks to violation
of the Reid criteria \cite{Reid(2)}, one-way steering has been demonstrated
in various works \cite{G(5)}, but these have mostly focused only on the
stationary regime.\newline
Besides being of fundamental interest, quantum steering has recently
attracted significant theoretical \cite{G(3)} and experimental \cite%
{Wollmann,G(4)} attention as an essential resource for a number of
applications, such as quantum key distribution \cite{Branciard}, secure
quantum teleportation \cite{Reid(3)} and randomness generation \cite{Bancal}.%
\newline
Motivated by the above mentioned achievements, we theoretically examine
optomechanical Gaussian quantum steering. For this, we consider two
spatially separated optomechanical Fabry-Perot cavities fed by broadband
two-mode squeezed light. In the resolved sideband regime with an adiabatic
elimination of the optical cavities modes, we investigate the Gaussian
steering and its asymmetry of two mixed mechanical modes, where a specific
attention is devoted to the dynamics of the one-way steerability. Moreover,
utilizing the two considered modes, we compare the Gaussian steering with
entanglement as two different aspects of inseparable quantum correlations.
In this way, we shall use the measure proposed recently by Kogias \textit{et
al} \cite{Kogias} as a quantifier of quantum steering for arbitrary
bipartite Gaussian states. To quantify entanglement, we will use the
Gaussian R\'{e}nyi-2 entropy \cite{AGS,AL}. Notice that in terms of the
difficulties in the creation of stationary entanglement and quantum
steering, the transient regime could be free from the decoherence issue and
the dissipation effects on one hand \cite{He and Ficek}; on the other hand,
the system under investigation could not be limited by the stability
requirements \cite{He and Ficek}. \newline
Finally, we note that in the past decade, optomechanical systems have been
attracted considerable interest (both theoretical and experimental) for
investigating various quantum phenomena \cite{Opto-prog,SGHofer}. Proposals
include, the creation of entangled states \cite{G(6)}, ground state optical
feedback cooling of the fundamental vibrational mode \cite{G(7)}, the
observation of quantum state transfer \cite{G(8)} and massive quantum
superpositions or so-called Schr\"{o}dinger cat states \cite{G(9)}.\newline
The remainder of this paper is organized as follows. In Sec. \ref{sec2}, we
present a detailed description of the optomechanical system under
investigation. We give the quantum Langevin equations governing the dynamics
of the mechanical and optical modes. The needed approximations to derive
closed analytical expression for the time-dependent covariance matrix of the
mechanical fluctuations are also discussed. In Sec. \ref{sec3}, using the
quantum steering formulation proposed in \cite{Kogias}, we study the
dynamics of Gaussian steering and its asymmetry for the two mechanical modes
taking into account thermal and squeezing effects. Also, we compare under
the same circumstance, the behavior of Gaussian steering of the two
considered modes with their corresponding entanglement. Finally, in Sec \ref%
{sec4} we draw our conclusions.

\section{ System and Hamiltonian \label{sec2}}

\subsection{ The model}

\begin{figure}[tbh]
\centerline{\includegraphics[width=9cm]{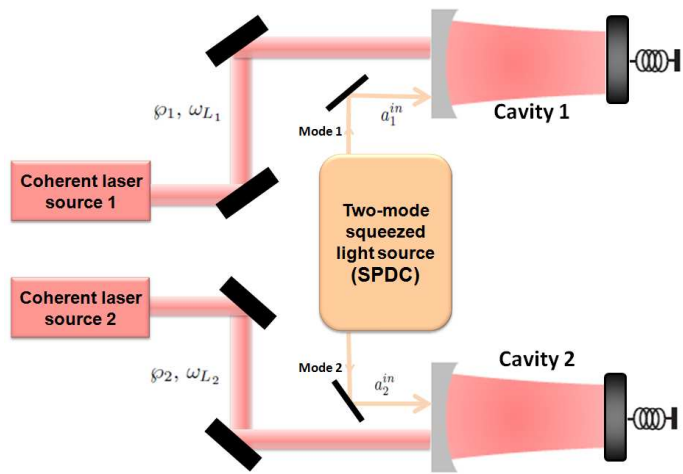}}
\caption{Schematics of two optomechanical Fabry-Perot cavities coupled to a
two-mode squeezed light from spontaneous parametric down-conversion (SPDC).
The $j^{th}$ cavity is pumped by a coherent laser field of power ${\wp }_{j}$
and frequency $\protect\omega _{L_{j}}$ for $j=1,2$. We will consider a
single mechanical mode of the $j^{th}$ movable mirror only, which can be
modeled as an harmonic oscillator with frequency $\protect\omega _{\protect%
\mu _{j}}$, a damping rate $\protect\gamma _{j}$ and an effective mass $m_{%
\protect\mu _{j}}$. $a_{j}^{in}$ is the $j^{th}$ noise operator
corresponding to the $j^{th}$ squeezed mode.}
\label{Fig.1}
\end{figure}
\noindent We consider two Fabry-Perot cavities in Fig. \ref{Fig.1}, where
each cavity is composed by two mirrors. The first one is fixed and partially
transmitting. The second is movable and perfectly reflecting. As depicted in
Fig. \ref{Fig.1}, the $j^{th}$ cavity is pumped by coherent laser field with
the input power ${\wp }_{j}$, phase $\varphi _{j}$ and frequency $\omega
_{L_{j}} $. In addition, the two cavities are also pumped by two-mode
squeezed light produced for example by spontaneous parametric
down-conversion source (SPDC) \cite{Burnham-Shih}. The first (respectively,
the second) squeezed mode is sent towards the first (second) cavity.
Finally, the $j^{th}$ movable mirror modeled as a quantum mechanical
harmonic oscillator \cite{M.K.Olsen} has an effective mass $m_{\mu _{j}}$, a
mechanical damping rate $\gamma _{j}$ and oscillates at frequency denoted by
$\omega _{\mu _{j}}$.

\subsection{ The Hamiltonian}

\noindent In a frame rotating at the frequency of the lasers, the
Hamiltonian of the two optomechanical cavities reads ( $\hbar =1$) \cite{Law}%
:
\begin{equation}
H=\sum_{j=1}^{2}\left[ \left( \omega _{c_{j}}-\omega _{L_{j}}\right)
a_{j}^{\dag }a_{j}+\omega _{\mu _{j}}b_{j}^{\dag }b_{j}+g_{j}a_{j}^{\dag
}a_{j}(b_{j}^{\dag }+b_{j})+\varepsilon _{j}(e^{i\varphi _{j}}a_{j}^{\dag
}+e^{-i\varphi _{j}}a_{j})\right] .  \label{E1}
\end{equation}%
where $b_{j},b_{j}^{\dag }$ are the annihilation and creation operators
associated with the mechanical mode describing the mirror $j$ (for $j=1,2$).
They satisfy the usual commutation relations $[b_{j},b_{k}^{\dag }]=\delta
_{jk}$ (for $j,k=1,2$). As we shall mainly be concerned in Sec. \ref{sec3}
with the quantum correlations between the mechanical modes, we will refer to
the first mode as Alice and to the second mode as Bob. Moreover, $a_{j}$ and
$a_{j}^{\dag }$ are the annihilation and creation operators of the $j^{th}$
optical cavity mode. They satisfy also the usual commutation $%
[a_{j},a_{k}^{\dag }]=\delta _{jk}$. The optomechanical single-photon
coupling rate $g_{j}$ between the $j^{th}$ mechanical mode and its
corresponding optical cavity mode is given by $g_{j}=\left( \omega
_{c_{j}}/l_{j}\right) \sqrt{\hbar /m_{\mu _{j}}\omega _{\mu _{j}}}$ where $%
l_{j}$ is the $j^{th}$ cavity length. The coupling strength between the $%
j^{th}$ external laser and its corresponding cavity field is defined by $%
\varepsilon _{j}=\sqrt{2\kappa _{j}{\wp }_{j}/\hbar \omega _{L_{j}}}$ , $%
\kappa _{j}$ being the energy decay rate of the $j^{th}$ cavity.

%%%%%%%%%%%%%%%%%%%%%%%%%%%%%%%%%%%%%%%%%%%%%%%%%%%%%%%%%

\subsection{Quantum Langevin equation}

In the Heisenberg picture, the dynamics of the $j^{th}$ mechanical and
optical mode variables is completely described by the following set of
nonlinear quantum Langevin equations:
\begin{eqnarray}
\partial _{t}b_{j} &=&-\left( \gamma _{j}/2+i\omega _{\mu _{j}}\right)
b_{j}-ig_{j}a_{j}^{\dag }a_{j}+\sqrt{\gamma _{j}}b_{j}^{in},  \label{E2} \\
\partial _{t}a_{j} &=&-\left( \kappa _{j}/2-i\Delta _{j}\right)
a_{j}-ig_{j}a_{j}(b_{j}^{\dag }+b_{j})-i\varepsilon _{j}e^{i\varphi _{j}}+%
\sqrt{\kappa _{j}}a_{j}^{in},  \label{E3}
\end{eqnarray}%
where $\Delta _{j}=\omega _{L_{j}}-\omega _{c_{j}}$ is the $j^{th}$ laser
detuning \cite{Marquardt} with $j=1,2$. Moreover, $b_{j}^{in}$ is the $%
j^{th} $ random Brownian operator, with zero mean value ($\langle
b_{j}^{in}\rangle= 0$), describing the coupling of the $j^{th}$ movable
mirror with its own environment. In general, $b_{j}^{in}$ is not $\delta$%
-correlated \cite{Giovanitti(2)}. However, quantum effects are reached only
using oscillators with a large mechanical quality factor $\mathcal{Q}%
=\omega_{\mu}/\gamma\gg 1$, which allows us to recover the Markovian
process. In this limit, we have the following nonzero time-domain
correlation functions \cite{Giovanitti(2),P.Zoller}:
\begin{eqnarray}
\langle b_{j}^{in\dag }(t)b_{j}^{in}(t^{\prime })\rangle &=&n_{\mathrm{th,}%
j}\delta (t-t^{\prime }),  \label{E4} \\
\langle b_{j}^{in}(t)b_{j}^{in\dag }(t^{\prime })\rangle &=&(n_{\mathrm{th,}%
j}+1)\delta (t-t^{\prime }),  \label{E5}
\end{eqnarray}%
where $n_{\mathrm{th,}j}=\left[ \exp (\hbar \omega _{\mu _{j}}/k_{B}T_{j})-1%
\right] ^{-1}$ is the mean thermal photon number, $T_{j}$ is the temperature
of the $j^{th}$ mirror environment and $k_{B}$ is the Boltzmann constant.
Another kind of noise affecting the system is the $j^{th}$ input squeezed
light noise operator $a_{j}^{in}$ with zero mean value ($\langle
a_{j}^{in}\rangle =0$). They have the following non-zero correlation
properties \cite{Gardiner,Mauro}:
\begin{eqnarray}
\langle \delta a_{j}^{in^{\dag }}(t)\delta a_{j}^{in}(t^{\prime })\rangle
&=&N\delta (t-t^{\prime })\text{ \ for \ \ }j=1,2,  \label{E6} \\
\langle \delta a_{j}^{in}(t)\delta a_{j}^{in^{\dag }}(t^{\prime })\rangle
&=&(N+1)\delta (t-t^{\prime })\text{ for \ \ }j=1,2,  \label{E7} \\
\langle \delta a_{j}^{in}(t)\delta a_{k}^{in}(t^{\prime })\rangle
&=&Me^{-i\omega _{\mu }(t+t^{\prime })}\delta (t-t^{\prime })\text{ \ for \ }%
j\neq k=1,2,  \label{E8} \\
\langle \delta a_{j}^{in^{\dag }}(t)\delta a_{k}^{in^{\dag }}(t^{\prime
})\rangle &=&Me^{i\omega _{\mu }(t+t^{\prime })}\delta (t-t^{\prime })\text{%
\ for \ }j\neq k=1,2,  \label{E9}
\end{eqnarray}%
where $N=\mathrm{sinh}^{\mathrm{2}}r$, $M=\mathrm{sinh}r\mathrm{cosh}r$, $r$
being the squeezing parameter (we have assumed that $\omega _{\mu
_{1}}=\omega _{\mu _{2}}=\omega _{\mu }$).

%%%%%%%%%%%%%%%%%%%%%%%%%%%%%%%%%%%%%%%%%%%%%%%%%%%%%%%%%%%%%%%

\subsection{ Linearization of quantum Langevin equations}

Due to the nonlinear nature of the radiation pressure, the coupled nonlinear
quantum Langevin equations (\ref{E2})-(\ref{E3}) are in general not solvable
analytically. To obtain analytical solution to these equations, we adopt the
linearization approach discussed in \cite{G.J.Milburn,Fabre}. We decompose
each operator ($a_{j}$ and $b_{j}$ for $j=1,2$) into two parts, i.e., sum of
its mean value and a small fluctuation with zero mean value. Thus, $\mathcal{%
O}_{j}=\langle \mathcal{O}_{j}\rangle +\delta \mathcal{O}_{j}=\mathcal{O}%
_{js}+\delta \mathcal{O}_{j}$ (with $\mathcal{O}_{j}\equiv a_{j},b_{j}$).
The mean values $b_{js}$ and $a_{js}$ are obtained by setting the time
derivatives to zero and factorizing the averages in Eqs. (\ref{E2}) and (\ref%
{E3}). Therefore, one gets
\begin{equation}
\langle a_{j}\rangle =a_{js}=\frac{-i\varepsilon _{j}e^{i\varphi _{j}}}{%
\kappa _{j}/2-i\Delta _{j}^{\prime }}\text{ \ \ \ and \ \ \ }\langle
b_{j}\rangle =b_{js}=\frac{-ig_{j}\left\vert a_{js}\right\vert ^{2}}{\gamma
_{j}/2+i\omega _{\mu _{j}}},\qquad  \label{E10}
\end{equation}%
where $\Delta _{j}^{\prime }$ $=$ $\Delta _{j}$ $-g_{j}(b_{js}^{\ast
}+b_{js})$ is the $j^{th}$ effective cavity detuning including the radiation
pressure effects \cite{Marquardt,Tombisi(2)}. To simplify further our
purpose, we assume that the double-cavity system is intensely driven ($%
\left\vert a_{js}\right\vert \gg 1$, for $j=1,2$). This assumption can be
realized considering lasers with a large input power $\wp _{j}$ \cite%
{Vitali(1)}. Consequently, the nonlinear terms $\delta a_{j}^{\dag }\delta
a_{j}$, $\delta a_{j}\delta b_{j}$ and $\delta a_{j}\delta b_{j}^{\dag }$
can be safely neglected. Hence, we obtain:
\begin{eqnarray}
\delta \dot{b}_{j} &=&-\left( \gamma _{j}/2+i\omega _{\mu _{j}}\right)
\delta b_{j}+G_{j}\left( \delta a_{j}-\delta a_{j}^{\dag }\right) +\sqrt{%
\gamma _{j}}b_{j}^{in},  \label{E11} \\
\delta \dot{a}_{j} &=&-\left( \kappa _{j}/2-i\Delta _{j}^{\prime }\right)
\delta a_{j}-G_{j}\left( \delta b_{j}^{\dag }+\delta b_{j}\right) +\sqrt{%
\kappa _{j}}\delta a_{j}^{in},  \label{E12}
\end{eqnarray}%
where $G_{j}$ $=g_{j}\left\vert a_{js}\right\vert $ is the $j^{th}$
light-enhanced optomechanical coupling in the linearized regime \cite%
{Marquardt}. It is given by:
\begin{equation}
G_{j}=\frac{\omega _{c_{j}}}{l_{j}}\sqrt{\frac{2\kappa _{j}\wp _{j}}{m_{\mu
_{j}}\omega _{\mu }\omega _{L_{j}}\left( \left( \frac{\kappa _{j}}{2}\right)
^{2}+\left( \Delta _{j}^{\prime }\right) ^{2}\right) }}.  \label{E12-}
\end{equation}
Notice that the Eqs. (\ref{E11}) and (\ref{E12}) have been obtained by
setting $a_{js}=-i\left\vert a_{js}\right\vert $ or equivalently to choose
the phase $\varphi _{j}$ of the $j^{th}$ input laser field to be $\varphi
_{j}=-\arctan (2\Delta _{j}^{\prime }/\kappa _{j})$. Now, we introduce the
operators $\delta \tilde{b}_{j}$ and $\delta \tilde{a}_{j}$ defined by $%
\delta b_{j}=\delta \tilde{b}_{j}e^{-i\omega _{\mu }t}$ and $\delta
a_{j}=\delta \tilde{a}_{j}e^{i\Delta _{j}^{\prime }t}$. Using the Eqs. (\ref%
{E11}) and (\ref{E12}), we obtain:
\begin{eqnarray}
\delta \dot{\tilde{b}}_{j} &=&-\frac{\gamma _{j}}{2}\delta \tilde{b}_{j}\
+G_{j}\left( \delta \tilde{a}_{j}e^{i\left( \Delta _{j}^{\prime }+\omega
_{\mu }\right) t}-\delta \tilde{a}_{j}^{\dag }e^{-i\left( \Delta
_{j}^{\prime }-\omega _{\mu }\right) t}\right) +\sqrt{\gamma _{j}}\tilde{b}%
_{j}^{in},  \label{E13} \\
\delta \dot{\tilde{a}}_{j} &=&-\frac{\kappa _{j}}{2}\delta \tilde{a}%
_{j}-G_{j}\left( \delta \tilde{b}_{j}e^{-i\left( \Delta _{j}^{\prime
}+\omega _{\mu }\right) t}+\delta \tilde{b}_{j}^{\dag }e^{-i\left( \Delta
_{j}^{\prime }-\omega _{\mu }\right) t}\right) +\sqrt{\kappa _{j}}\delta
\tilde{a}_{j}^{in}.  \label{E14}
\end{eqnarray}%
Next, we assume that the two cavities are driven at \textit{the red sideband}
($\Delta _{j}^{\prime }=-\omega _{\mu }$ for $j=1,2$) which corresponds to
the quantum states transfer regime \cite{G(8),Pinard}. We note also that, in
the resolved-sideband regime where the mechanical frequency $\omega _{\mu }$
of the movable mirror is larger than the $j^{th}$ cavity decay rate $\kappa
_{j} $ ($\omega _{\mu }\gg \kappa _{1}$, $\kappa _{2}$), one can use the
rotating wave approximation (RWA) \cite{Marquardt,Clerk(2)}, allowing us to
ignore terms rotating at $\pm 2\omega _{\mu }$ in equations (\ref{E13}) and (%
\ref{E14}). Then, one gets
\begin{eqnarray}
\quad \delta \dot{\tilde{b}}_{j} &=&-\frac{\gamma _{j}}{2}\delta \tilde{b}%
_{j}+G_{j}\delta \tilde{a}_{j}+\sqrt{\gamma _{j}}\tilde{b}_{j}^{in},
\label{E15} \\
\delta \dot{\tilde{a}}_{j} &=&-\frac{\kappa _{j}}{2}\delta \tilde{a}%
_{j}-G_{j}\delta \tilde{b}_{j}+\sqrt{\kappa _{j}}\delta \tilde{a}_{j}^{in}.
\label{E16}
\end{eqnarray}

%%%%%%%%%%%%%%%%%%%%%%%%%%%%%%%%%%%%%%%%%%%%%%%%%%%%%%%%%%%%%%%%%%%%

\subsection{The adiabatic elimination of the optical modes}

%%%%%%%%%%%%%%%%%%%%%%%%%%%%%%%%%%%%%%%%%%%%%%%%%%%%%%%%%%%%%%%%%
Being interested only in the quantum correlations between two mechanical
modes, the optimal regime for quantum fluctuations transfer from the
two-mode squeezed light to the two movable mirrors is achieved when the
optical cavities modes adiabatically follow the mechanical modes, which
corresponds to the situation where the mirrors have a large mechanical
quality factor and weak effective optomechanical coupling ($\kappa _{j}\gg
G_{j},$ $\gamma _{j}$) \cite{Pinard(2)}. In this way, inserting the steady
state solution of (\ref{E16}) into (\ref{E15}), we obtain a simple
description for the two mechanical modes. Then, the $j^{th}$ mirror dynamics
reduces to:
\begin{equation}
\delta \dot{\tilde{b}}_{j}=-\frac{\Gamma _{j}}{2}\delta \tilde{b}_{j}+\sqrt{%
\gamma _{j}}\tilde{b}_{j}^{in}+\sqrt{\Gamma _{\mathrm{a}_{j}}}\delta \tilde{a%
}_{j}^{in}=-\frac{\Gamma _{j}}{2}\delta \tilde{b}_{j}+\tilde{F}_{j}^{in},
\label{E17}
\end{equation}%
where $\Gamma _{\mathrm{a}_{j}}=4G_{j}^{2}/\kappa _{j}$ is the effective
relaxation rate induced by radiation pressure \cite{Karrai}, $\Gamma _{j}=$ $%
\Gamma _{\mathrm{a}_{j}}+\gamma _{j}$ and $\tilde{F}_{j}^{in}=\sqrt{\gamma
_{j}}\tilde{b}_{j}^{in}+\sqrt{\Gamma _{\mathrm{a}_{j}}}\delta \tilde{a}%
_{j}^{in}$. Defining the mechanical fluctuation quadratures, and their
corresponding Hermitian input noise operators:
\begin{eqnarray}
\delta \tilde{q}_{j} &=&(\delta \tilde{b}_{j}^{\dagger }+\delta \tilde{b}%
_{j})/\sqrt{2},\quad \delta \tilde{p}_{j}=i(\delta \tilde{b}_{j}^{\dagger
}-\delta \tilde{b}_{j})/\sqrt{2},  \label{E 18 -2} \\
\tilde{F}_{q_{j}}^{in} &=&(\tilde{F}_{j}^{in,\dag }+\tilde{F}_{j}^{in})/%
\sqrt{2},\qquad \tilde{F}_{p_{j}}^{in}=i(\tilde{F}_{j}^{in,\dag }-\tilde{F}%
_{j}^{in})/\sqrt{2},  \label{E 18 -4}
\end{eqnarray}%
the linearized quantum Langevin equations can be written in the following
compact matrix form \cite{Eisert(1)}:
\begin{equation}
\dot{u}(t)=Su(t)+n(t),  \label{E20}
\end{equation}%
where $S=\mathrm{diag}(-\frac{\Gamma _{1}}{2},-\frac{\Gamma _{1}}{2},-\frac{%
\Gamma _{2}}{2},-\frac{\Gamma _{2}}{2})$, $u(t)^{\mathrm{T}}=(\delta \tilde{q%
}_{1},\delta \tilde{p}_{1},\delta \tilde{q}_{2},\delta \tilde{p}_{2})$ and $%
n(t)^{\mathrm{T}}=(\tilde{F}_{q_{1}}^{in},\tilde{F}_{p_{1}}^{in},\tilde{F}%
_{q_{2}}^{in},\tilde{F}_{p_{2}}^{in})$. The system is stable only if the
real parts of all the eigenvalues of the drift matrix $S$ are negative,
which is fully verified according to the form of the matrix $S$. Such
stability is guaranteed by the fact that both pumps drive the resonators on
the red sideband. Therefore, the use of the Routh-Hurwitz criterion \cite{R
and H} is without interest. Nonetheless since we have linearized the
dynamics and the noises are zero-mean quantum Gaussian noises, fluctuations
in the stable regime will also evolve to an asymptotic zero-mean Gaussian
state. It follows that the state of the system is completely described by
the correlation matrix $V(t)$ of elements:
\begin{equation}
V_{ii^{\prime }}(t)=\frac{1}{2}(\langle u_{i}(t)u_{i^{\prime
}}(t)+u_{i^{\prime }}(t)u_{i}(t)\rangle ).  \label{E21}
\end{equation}%
Using Eqs. (\ref{E20}) and (\ref{E21}), the matrix $V(t)$ satisfies the
following evolution equation \cite{Eisert(1)}:
\begin{equation}
\frac{d}{dt}V(t)=SV(t)+V(t)S^{\mathrm{T}}+D,  \label{E22}
\end{equation}%
where $D$ is the noise correlation matrix defined by $D_{kk^{\prime }}\delta
(t-t^{\prime })=(\langle n_{k}(t)n_{k^{\prime }}(t^{\prime })+n_{k^{\prime
}}(t^{\prime })n_{k}(t)\rangle )/2$. Utilizing the correlation properties of
the noise operators given by the set of equations [(\ref{E4})-(\ref{E9})],
we obtain:
\begin{equation}
D=\left(
\begin{array}{cccc}
\ D_{11}\ \  & 0 & D_{13}\  & 0 \\
0 & \ \ D_{22}\ \  & 0 & D_{24}\  \\
D_{13}\  & 0 & \ D_{33}\ \  & 0 \\
0 & D_{24}\  & 0 & D_{44}\ \
\end{array}%
\right) ,  \label{E23}
\end{equation}%
where $D_{11}\ \ =D_{22}=\Gamma _{\mathrm{a}_{1}}\left( N+1/2\right) +\gamma
_{1}\left( n_{\mathrm{th,}1}+1/2\right) $, $D_{33}=D_{44}=\Gamma _{\mathrm{a}%
_{2}}\left( N+1/2\right) +\gamma _{2}\left( n_{\mathrm{th,}2}+1/2\right) $
and $D_{13}\ =-D_{24}=M\sqrt{\Gamma _{\mathrm{a}_{1}}\Gamma _{\mathrm{a}_{2}}%
}$. The equation (\ref{E22}) is an ordinary linear differential equation and
can be solved straightforwardly. The corresponding solution can be written
as:
\begin{equation}
V(t)=\left(
\begin{array}{cccc}
v_{11}(t) & 0 & v_{13}(t) & 0 \\
0 & v_{22}(t) & 0 & v_{24}(t) \\
v_{13}(t) & 0 & v_{33}(t) & 0 \\
0 & v_{24}(t) & 0 & v_{44}(t)%
\end{array}%
\right) \equiv \left(
\begin{array}{cc}
V_{1}(t) & V_{3}(t) \\
V_{3}^{\mathrm{T}}(t) & V_{2}(t)%
\end{array}%
\right) ,  \label{E24}
\end{equation}%
with $V_{1}(t)=\mathrm{diag}(v_{11}(t),v_{22}(t))$, $V_{2}(t)=\mathrm{diag}%
(v_{33}(t),v_{44}(t))$ and $V_{3}(t)=\mathrm{diag}(v_{13}(t),v_{24}(t))$. We
note that $V(t)$ is a real, symmetric and positive definite matrix. The $%
2\times 2$ matrices $V_{1}(t)$ and $V_{2}(t)$ represent the first and second
mechanical mode respectively, while the correlations between them are
described by $V_{3}(t)$. Considering identical damping rates ($\gamma
_{1}=\gamma _{2}=\gamma $), the explicit expressions of the covariance
matrix elements are given by:
\begin{eqnarray}
v_{11}(t) &=&v_{22}(t)=\frac{(2N+1)\mathcal{C}_{1}+2n_{\mathrm{th,1}}+1}{2(%
\mathcal{C}_{1}+1)}+\frac{(-2N+1)\mathcal{C}_{1}-2n_{\mathrm{th,1}}+1}{2(%
\mathcal{C}_{1}+1)}e^{-\gamma (\mathcal{C}_{1}+1)t},  \label{E25} \\
v_{33}(t) &=&v_{44}(t)=\frac{(2N+1)\mathcal{C}_{2}+2n_{\mathrm{th,2}}+1}{2(%
\mathcal{C}_{2}+1)}+\frac{(-2N+1)\mathcal{C}_{2}-2n_{\mathrm{th,2}}+1}{2(%
\mathcal{C}_{2}+1)}e^{-\gamma (\mathcal{C}_{2}+1)t},  \label{E26} \\
v_{13}(t) &=&-v_{24}(t)=\frac{2M\sqrt{\mathcal{C}_{1}\mathcal{C}_{2}}}{%
\mathcal{C}_{1}+\mathcal{C}_{2}+2}\left( 1-e^{-\frac{\gamma }{2}(\mathcal{C}%
_{1}+\mathcal{C}_{2}+2)t}\right) ,  \label{E27}
\end{eqnarray}%
where $\mathcal{C}_{j}$ is the $j^{th}$ optomechanical cooperativity \cite%
{Kampel}:
\begin{equation}
\mathcal{C}_{j}=\Gamma _{\mathrm{a}_{j}}/\gamma =4G_{j}^{2}/\gamma \kappa
_{j}=\frac{8\omega _{c_{j}}^{2}}{\gamma m_{\mu _{j}}\omega _{\mu }\omega
_{L_{j}}l_{j}^{2}}\frac{\ \wp _{j}}{\left[ \left( \frac{\kappa _{j}}{2}%
\right) ^{2}+\omega _{\mu }^{2}\right] }.  \label{E28}
\end{equation}
In the strong optomechanical coupling regime, where $\mathcal{C}_{1,2}\gg 1$
(in the limit of strong coupling $\mathcal{C}_{1,2}\rightarrow10^{6}$ \cite%
{Wang and Clerk}) and longer time, $v_{11}(t)$, $v_{33}(t)$ and $v_{13}(t)$
reduce respectively to $v_{11}(t)=v_{33}(t)= \frac{1}{2}\mathrm{cosh} (2r)$
and $v_{13}(t)= \frac{1}{2}\mathrm{sinh} (2r)$, meaning that quantum
correlations can be governed only by the squeezing degree $r$. Moreover,
when either $\mathcal{C}_{1}=0$, $\mathcal{C}_{2}=0$ or $r=0$, we have $%
v_{13}(t)=v_{24}(t)=0$ or equivalently $\det V _{3}=0$, which corresponds to
the Gaussian product states \cite{G(1)}, so that the two modes $A$ and $B$
remain separable and consequently, they would be non-steerable in any
direction \cite{Kogias}. This is a consequence of the fact that $\det V
_{3}<0$ is a necessary condition for a two-mode Gaussian state to be
entangled \cite{Simon}. Therefore, non zero optomechanical coupling and non
zero squeezing are necessary conditions to correlate the two separated modes
$A$ and $B$. Finally, from Eqs. [(\ref{E25})-(\ref{E27})], it is not
difficult to show that when $t\rightarrow\infty$ which corresponds to the
stationary regime, $v_{11}(\infty)$, $v_{33}(\infty)$ and $v_{13}(\infty)$
coincide respectively with Eqs. (21), (22) and (23) in Ref \cite{Sete}.
%%%%%%%%%%%%%%%%%%%%%%%%%%%%%%%%%%%%%%%%%%%%%%%%%%%%%%%%%%%%%%%%%%%%%%%%%%%%%%%%%%%%%%%%%

\section{Gaussian quantum steering and its asymmetry \label{sec3}}

Now, we are in position to study the dynamics of Gaussian quantum steering
and its asymmetry between the two mechanical modes $A$ and $B$. To quantify
how much a bipartite Gaussian state with covariance matrix $V(t)$ is
steerable, we use the compact formula which has been proposed recently in
Ref \cite{Kogias}. Let us start by giving the definition of quantum
steerability. Following \cite{Wiseman,Kogias}, a bipartite state $\varrho
_{AB}$ is steerable from $A$ to $B$ (i.e., Alice can steer Bob's states)
after accomplishing a set of measurements $\mathcal{M}_{A}$ on Alice's side
iff it is not possible for every pair of local observables $R_{A}\in
\mathcal{M}_{A}$ on $A$ and $R_{B}$ (arbitrary) on $B$, with respective
outcomes $r_{A}$ and $r_{B}$, to express the joint probability as $%
P(r_{A},r_{B}|R_{A},R_{B},\varrho _{AB})=\sum_{\lambda }\mathcal{P}_{\lambda
}\mathcal{P}(r_{A},R_{A}|\lambda )P(r_{B},R_{B}|\varrho _{\lambda })$ \cite%
{Wiseman}. This means that at least one measurement pair $R_{A}$ and $R_{B}$
must violate this expression when $\mathcal{P}_{\lambda }$ is fixed across
all measurements \cite{Wiseman,Kogias}. Here $\mathcal{P}_{\lambda }$ and $%
\mathcal{P}(r_{A},R_{A}|\lambda )$ are\ probability distributions and $%
P(r_{B},R_{B}|\varrho _{\lambda })$\ is the conditional probability
distribution associated to the extra condition of being evaluated on the
state $\varrho _{\lambda }$.\newline
A bipartite system of \textit{two-mode Gaussian state} $\varrho _{AB}$ with
covariance matrix $V$ (Eq. (\ref{E24})) is $A\rightarrow B$ steerable by
Alice's Gaussian measurements, iff the following condition is violated \cite%
{Wiseman}:
\begin{equation}
V+i(0_{A}\oplus \Omega _{B})\geqslant 0,  \label{S1-}
\end{equation}%
where $0_{A}$ is a $2\times 2$ null matrix and $\Omega _{B}$ $=\left(
\begin{array}{cc}
0 & 1 \\
-1 & 0%
\end{array}%
\right) $ is the $B$-mode symplectic matrix \cite{Kogias}. Henceforth, a
violation of the condition (\ref{S1-}) is necessary and sufficient for the
Gaussian $A\rightarrow B$ steerability \cite{Kogias}. \newline
A computable measure to quantify how much a bipartite \textit{two-mode}
Gaussian state with covariance matrix $V$ (\ref{E24}) is steerable by
Gaussian measurements on Alice's side, is given by \cite{Kogias}:
\begin{equation}
\mathcal{G}^{A\rightarrow B}(V):=\max \{0,-\ln (\bar{\nu}^{B})\},
\label{S2-}
\end{equation}
with $\bar{\nu}^{B}=\sqrt{\det M^{B}}$ the symplectic eigenvalue of the
matrix $M^{B}$ written as $M^{B}=V_{2}-V_{3}^{\mathrm{T}}V_{1}^{-1}V_{3}$,
where the $2\times 2$ matrices $V_{1}$, $V_{2}$ and $V_{3} $ are defined by
Eq. (\ref{E24}). \newline
The Gaussian quantum steering $\mathcal{G}^{A\rightarrow B}$ vanishes when
the state described by the covariance matrix $V$ (\ref{E24}) is nonsteerable
by Alice's measurements, and it generally quantifies the amount by which the
condition (\ref{S1-}) fails to be fulfilled \cite{Kogias}. With quadratures
given by [(\ref{E 18 -2})-(\ref{E 18 -4})] and the covariance matrix (\ref%
{E24}) expressed in the ordered basis ($\delta \tilde{q}_{1},\delta \tilde{p}%
_{1},\delta \tilde{q}_{2},\delta \tilde{p}_{2}$), the Gaussian steerability $%
A\rightarrow B$ given by Eq. (\ref{S2-}) takes the following simple form\cite%
{Kogias}:
\begin{equation}
\mathcal{G}^{A\rightarrow B}=\max \left[ 0,\frac{1}{2}\ln \frac{\det V_{1}(t)%
}{4\det V(t)}\right] =\max \left[ 0,\text{ }-\ln 2\left( v_{33}(t)-\frac{%
(v_{13}(t))^{2}}{v_{11}(t)}\right) \right] ,  \label{E30}
\end{equation}%
where $v_{11}(t),v_{33}(t)$ and $v_{13}(t)$ are explicitly given by Eqs. [(%
\ref{E25})-(\ref{E27})]. Similarly, a corresponding measure of the Gaussian $%
B\rightarrow A$ steerability can be obtained by swapping the roles of $A$
and $B$ in (\ref{E30}). One gets:
\begin{equation}
\mathcal{G}^{B\rightarrow A}=\max \left[ 0,\frac{1}{2}\ln \frac{\det V_{2}(t)%
}{4\det V(t)}\right] =\max \left[ 0,\text{ }-\ln 2\left( v_{11}(t)-\frac{%
(v_{13}(t))^{2}}{v_{33}(t)}\right) \right] .  \label{E31}
\end{equation}
The explicit analytical expressions of $\mathcal{G}^{A\rightarrow B}$ and $%
\mathcal{G}^{B\rightarrow A}$ are too cumbersome and will not be reported
here. It is well known that quantum entanglement is a symmetric property
shared between two systems $A$ and $B$ without specification of direction,
i.e., if $A$ is entangled with $B$, $B$ is necessarily entangled with $A$.
However, quantum steering is an asymmetric property ,i.e., a quantum state
may be steerable from Alice to Bob, but not vice versa \cite{Kogias}. Thus,
we shall consider three cases: ($i$) $\mathcal{G}^{A\rightarrow B}=\mathcal{G%
}^{B\rightarrow A}=0$ as no-way steering, ($ii$) $\mathcal{G}^{A\rightarrow
B}>0$ and $\mathcal{G}^{B\rightarrow A}=0$ or $\mathcal{G}^{A\rightarrow
B}=0 $ and $\mathcal{G}^{B\rightarrow A}>0$ as one-way steering, and finally
($iii $) $\mathcal{G}^{A\rightarrow B}>0$ and $\mathcal{G}^{B\rightarrow A}>0
$ as two-way steering. In order to check how asymmetric can the steerability
be between the mechanical modes $A$ and $B$, we use the Gaussian steering
asymmetry $\mathcal{G}_{AB}^{\Delta }$ defined as \cite{Kogias}:
\begin{equation}
\mathcal{G}_{AB}^{\Delta }=\left\vert \mathcal{G}^{A\rightarrow B}-\mathcal{G%
}^{B\rightarrow A}\right\vert .  \label{E32}
\end{equation}%
On the other hand, to compare between quantum steering and entanglement as
two different aspects of inseparable quantum correlations, it is more
convenient to plot them simultaneously under the same circumstances. To
accomplish this, we use the Gaussian R\'{e}nyi-$2$ entropy \cite{AGS,AL} as
an appropriate measure to quantify entanglement between the two modes $A$
and $B$ \cite{Kogias}.\newline
In quantum information theory, an interesting family of additive entropies
is represented by R\'{e}nyi-$\alpha $ entropies \cite{Renyi} defined by \cite%
{AGS}:
\begin{equation}
\mathcal{S}_{\alpha }\mathcal{(\varrho )=(}1-\alpha \mathcal{)}^{-1}\ln
\mathrm{Tr}\left( \varrho ^{\alpha }\right) .  \label{Renyi-alpha}
\end{equation}
In particular, when $\alpha \rightarrow 1,$ the entropies given by Eq. (\ref%
{Renyi-alpha}) reduce to the von Neumann entropy \newline
$\mathcal{S(\varrho )=}-\mathrm{Tr}\left( \varrho \ln \varrho \right) $ \cite%
{AGS}, which quantifies the degree of information contained in a quantum
state $\mathcal{\varrho }.$ While, for $\alpha =2$, we obtain the Gaussian R%
\'{e}nyi-2 entropy defined by \cite{AGS}:
\begin{equation}
\mathcal{S}_{2}\mathcal{(\varrho )=}-\ln \mathrm{Tr}\left( \varrho
^{2}\right) .  \label{Renyi-2}
\end{equation}
It has been shown in \cite{AGS}, that R\'{e}nyi-2 entropy provides a natural
measure of information for any multimode Gaussian state of quantum harmonic
systems. Importantly, it has been demonstrated also in \cite{AGS} that for
all Gaussian states, R\'{e}nyi-2 entropy satisfies the strong subadditivity
inequality, i.e., $\mathcal{S}_{2}\left( \mathcal{\varrho }_{AB}\right) +%
\mathcal{S}_{2}\left( \mathcal{\varrho }_{BC}\right) \geqslant \mathcal{S}%
_{2}\left( \mathcal{\varrho }_{ABC}\right) +\mathcal{S}_{2}\left( \mathcal{%
\varrho }_{B}\right) $, which made it possible to define measures of
Gaussian R\'{e}nyi-2 entanglement \cite{AL,Wolf and Cirac} and discord-like
quantum correlations \cite{G(1),ZHM}. \newline
For generally mixed two-mode Gaussian states $\mathcal{\varrho }_{AB}$, the R%
\'{e}nyi-2 entanglement measure $\mathcal{E}_{2}\mathcal{(\varrho }_{A:B}%
\mathcal{)\equiv E}_{2}$, defined by Eq. (\ref{Renyi-2}), admits an unclosed
cumbersome formula which will not be reported here \cite{AGS, AL}. However,
for relevant subclasses of states including symmetric states \cite{Giedke
and Wolf}, squeezed thermal states \cite{G(1)}, and so-called GLEMS-Gaussian
states of partial minimum uncertainty \cite{AL}, closed formulas of Gaussian
R\'{e}nyi-2 entanglement have been found \cite{AL}. The covariance matrix $%
V(t)$ (\ref{E24}) is in the so-called standard form \cite{Simon} and
characterized by $v_{13}(t)=-v_{24}(t)$ (\ref{E27}) which corresponds to the
\textit{squeezed thermal states STS} \cite{G(1)}. Therefore, the Gaussian R%
\'{e}nyi-2 entanglement measure $\mathcal{E}_{2}$, admits the following
expression \cite{AGS,AL}:
\begin{equation}
\mathcal{E}_{2}=\frac{1}{2}\ln \left[ h(s,d,g)\right] ,  \label{E33}
\end{equation}%
with:
\begin{equation}
h(s,d,g)=\left\{
\begin{array}{c}
1\text{ \ \ \ \ \ \ \ \ \ \ \ \ \ \ \ \ \ \ \ \ \ \ \ \ \ \ \ \ \ \ \ \ \ \
\ \ \ \ iff\ \ \ \ \ \ \ \ \ }4g\geqslant 4s-1, \\
\left[ \frac{(4g+1)s-\sqrt{\left[ (4g-1)^{2}-16d^{2}\right] \left[
s^{2}-d^{2}-g\right] }}{4(d^{2}+g)}\right] ^{2}\text{ \ iff \ \ \ \ }%
4|d|+1\leq 4g<4s-1,%
\end{array}%
\right.  \label{E-33}
\end{equation}%
where $s=\frac{1}{2}(v_{11}(t)+v_{33}(t)),$ $d=\frac{1}{2}%
(v_{11}(t)-v_{33}(t))$ and $g=\left( v_{11}(t)v_{33}(t)-v_{13}^{2}(t)\right)
$.
\begin{figure}[t]
\centerline{\includegraphics[width=0.5\columnwidth,height=4.5cm]{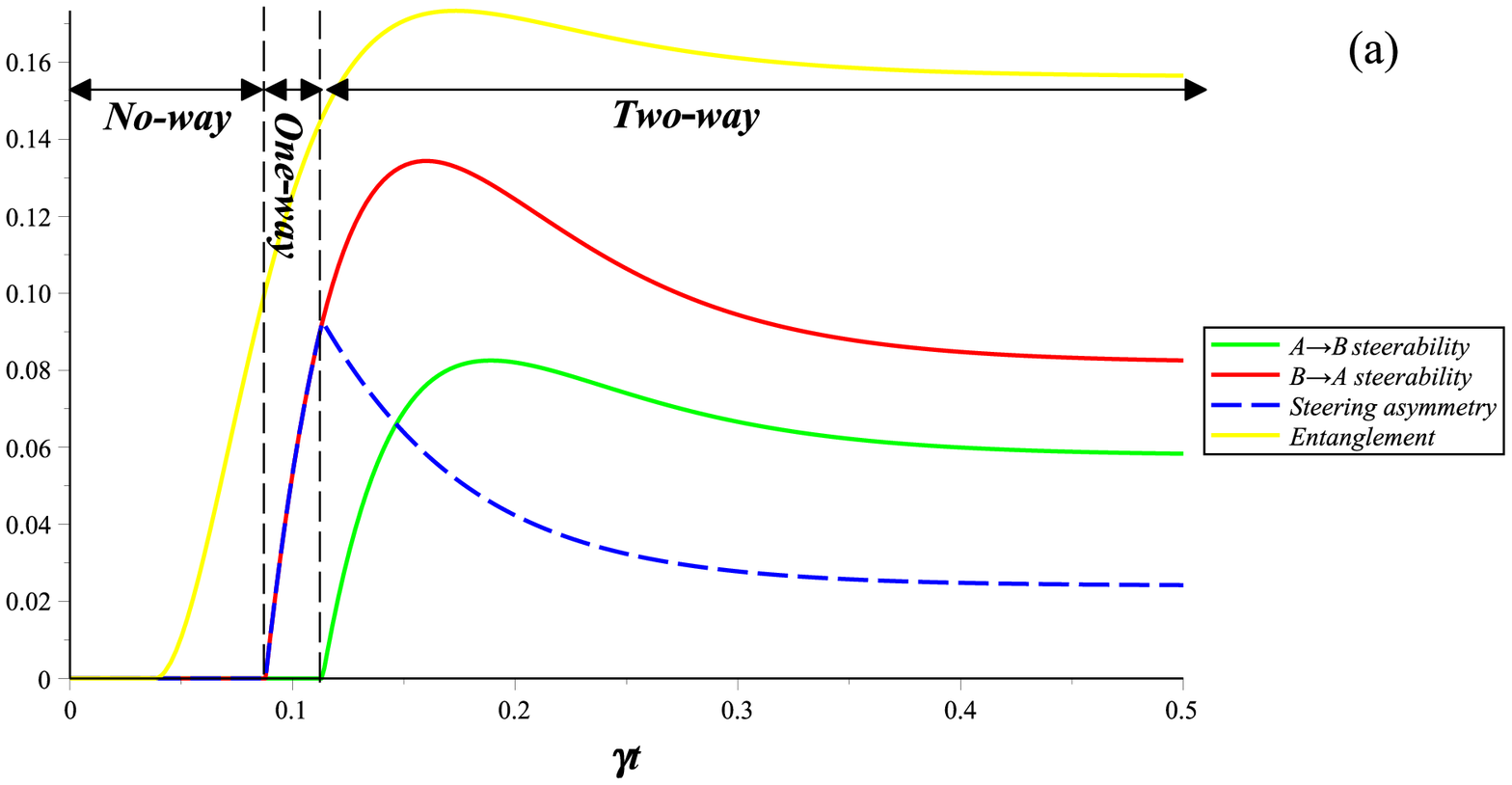}%
\includegraphics[width=0.5\columnwidth,height=4.5cm]{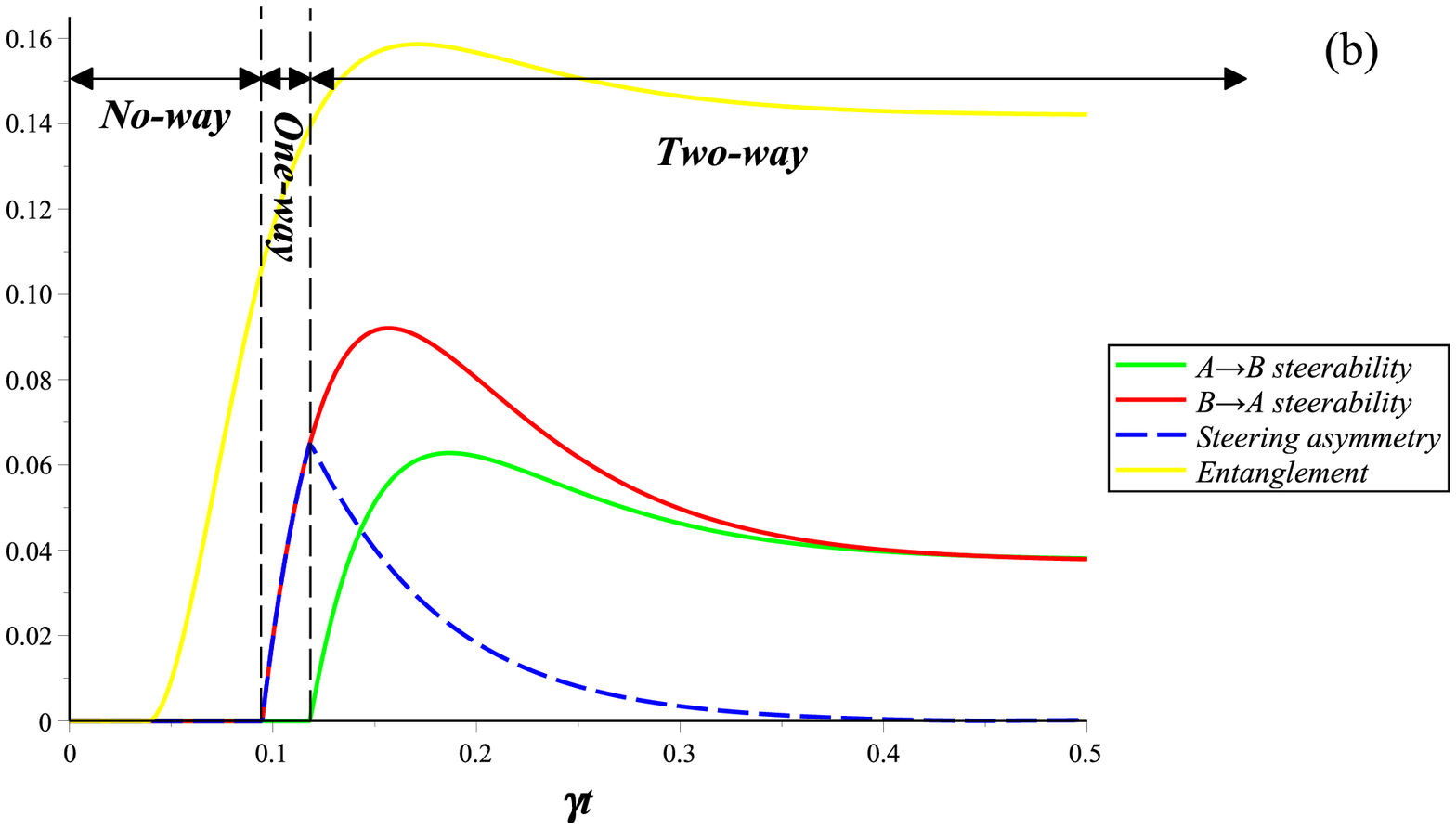}} %
\centerline{\includegraphics[width=0.5\columnwidth,height=4.5cm]{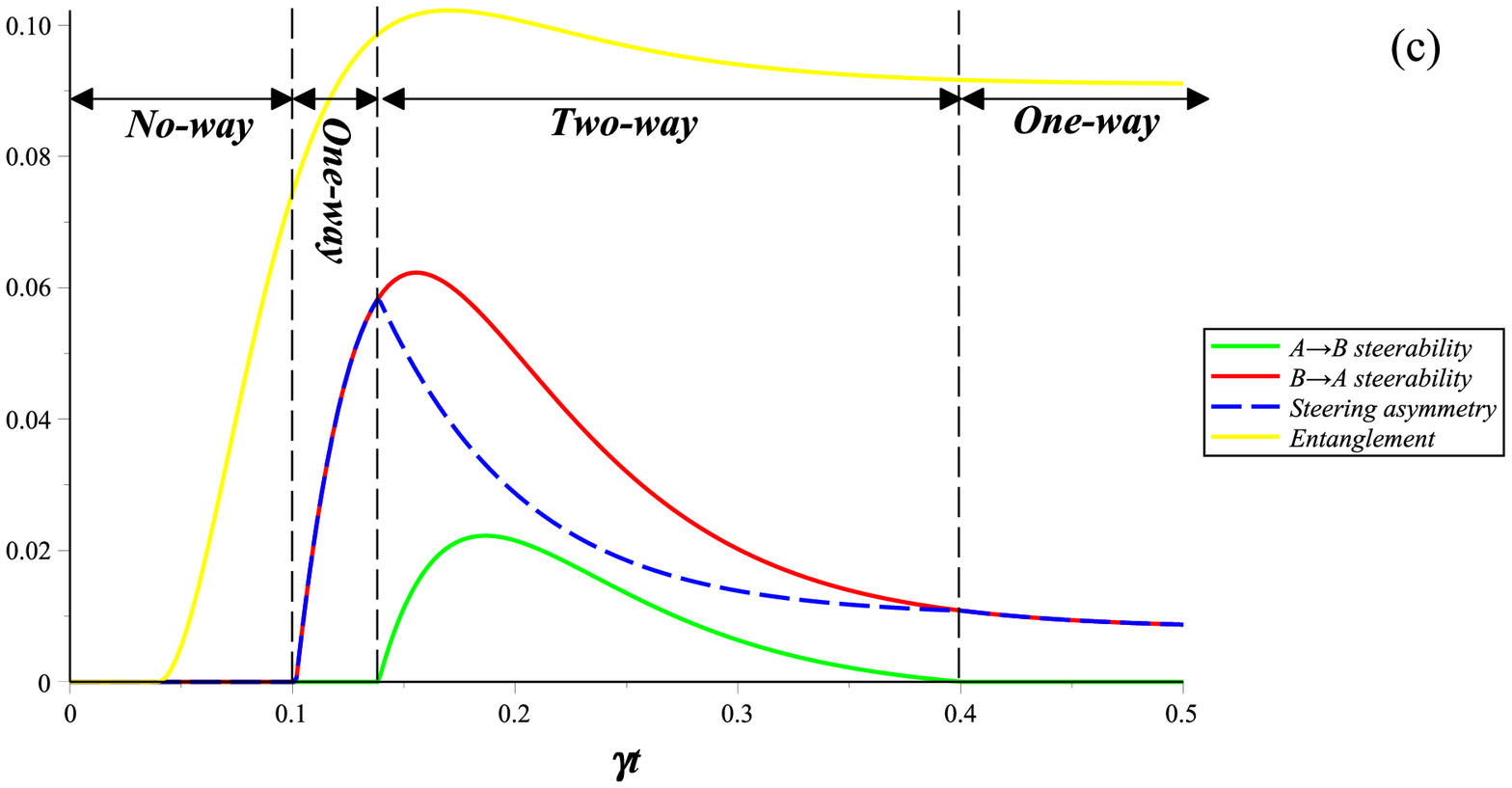}%
\includegraphics[width=0.5\columnwidth,height=4.5cm]{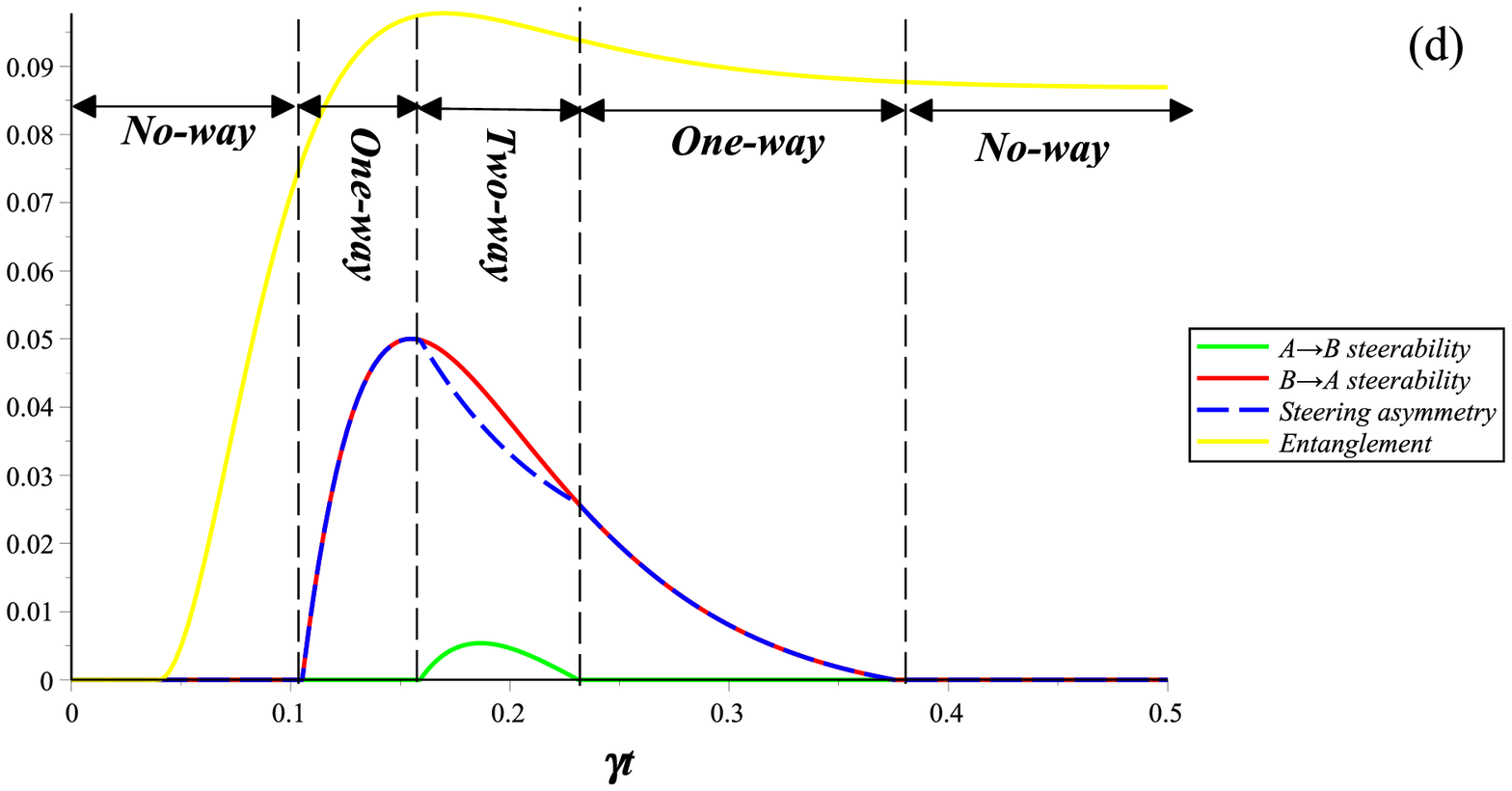}}
\caption{Plot of the Gaussian steering $\mathcal{G}^{A\rightarrow B}$ (green
solid line), $\mathcal{G}^{B\rightarrow A}$ (red solid line), the steering
asymmetry $\mathcal{G}_{AB}^{\Delta }$ (blue dashed line) and entanglement $%
\mathcal{E}_{2}$ (yellow solid line) of the two mechanical modes $A$ and $B$
as a function of the scaled time $\protect\gamma t$ for $\mathcal{C}_{1}=15$%
, $\mathcal{C}_{2}=35$ and $r=1$. The mean thermal photons numbers $n_{%
\mathrm{th},1}$ and $n_{\mathrm{th},2}$ are fixed as : $n_{\mathrm{th}%
,1}=0.5 $ and ${n_{\mathrm{th},2}=1}$ (panel (a)), $n_{\mathrm{th},1}=1$ and
$n_{\mathrm{th},2}=0.5$ (panel (b)), $n_{\mathrm{th},1}=1$ and $n_{\mathrm{th%
},2}=1.2$ (panel (c)) and $n_{\mathrm{th},1}=1$ and $n_{\mathrm{th},2}=1.5$
(panel (d)). Obviously, panels (a) and (b) show that when interchanging the
values of $n_{\mathrm{th,}1}$ and $n_{\mathrm{th,}2}$, the Gaussian R\'{e}%
nyi-2 entanglement $\mathcal{E}_{2}$ is insensitive to this operation
(unlike the steerabilities $\mathcal{G}^{A\rightarrow B}$, $\mathcal{G}%
^{B\rightarrow A}$), which means that the Gaussian R\'{e}nyi-2 entanglement
is unable to detect the asymmetry shown by the steering criteria (see Eqs. (%
\protect\ref{E30})-(\protect\ref{E31})). This figure shows also that
steerable states are strictly inseparable but not necessarily vice versa.}
\label{Fig.2}
\end{figure}
\noindent The expressions of $\mathcal{G}^{A\rightarrow B}$, $\mathcal{G}%
^{B\rightarrow A}$, $\mathcal{G}_{AB}^{\Delta }$ and $\mathcal{E}_{2}$
involve the covariance matrix elements (\ref{E24}) which are expressed in
terms of the squeezing parameter $r$, the $j^{th}$ optomechanical
cooperativity $\mathcal{C}_{j}$ and the $j^{th}$ mean thermal photons number
$n_{\mathrm{th,}j}$. In what follows, we shall consider the case where $n_{%
\mathrm{th,}1}\neq n_{\mathrm{th,}2}$ and $\mathcal{C}_{1}\neq \mathcal{C}%
_{2}$ so that the system is not symmetric by swapping the first and the
second mode, which is a crucial condition to ensure the Gaussian steering
asymmetry. In our simulations, the system parameters have been taken from
\cite{Groblacher}. The movable mirrors having the mass $\mu _{1,2}=145~%
\mathrm{ng}$ and oscillate at frequency $\omega _{\mu _{1,2}}=2\pi \times
947\times 10^{3}~\mathrm{Hz}$ with a mechanical damping rate $\gamma
_{1,2}=2\pi \times 140~\mathrm{Hz}$. The two cavities have length $%
l_{1,2}=25~\mathrm{\ mm}$, wave length $\lambda _{1,2}=1064~\mathrm{nm}$,
decay rate $\kappa _{1,2}=2\pi \times 215\times 10^{3}~\mathrm{Hz}$,
frequency $\omega _{c_{1,2}}=2\pi \times 5.26\times 10^{14}~\mathrm{Hz}$ and
pumped by laser fields of frequency $\omega _{L_{1,2}}=2\pi \times
2.82\times 10^{14}$ $\mathrm{Hz}$. For the powers of the coherent laser
sources, we take $\wp _{1}=5~\mathrm{mW}$ and $\wp _{2}=11~\mathrm{mW}$ \cite%
{Groblacher}. Next, using the explicit expression of the dimensionless $%
j^{th}$ optomechanical cooperativity $\mathcal{C}_{j}$ given by Eq. (\ref%
{E28}), one has $\mathcal{C}_{1}$ $\simeq $ $35$ and $\mathcal{C}_{2}$ $%
\simeq 15$. Taking the above parameters into account, we find the following
sequence of inequalities:
\begin{equation}
\omega _{\mu }\gg \kappa \gg G_{j} ,  \label{EE}
\end{equation}
where the parameter $G_{j}$ (for $j=1,2$) is given by Eq. (\ref{E12-}). So,
in accordance with \cite{SGHofer,Marquardt,Vitali(1),Clerk(2),Sun}, the
condition $\omega _{\mu }\gg \kappa $ justifies the use of the rotating wave
approximation. Meanwhile, $\kappa \gg G_{j} $ which is the condition of the
weak-coupling regime, allows us the adiabatic elimination of the optical
cavities modes \cite{SGHofer,Vitali(1),Pinard(2),Sun}. Concerning the
environmental parameters (the squeezing parameter $r$ and the thermal
occupations $n_{\mathrm{th},1}$ and $n_{\mathrm{th},2}$), we have chosen
them of the same order of magnitude as those used in \cite{GSAgarawal}.
\newline
Fixing the squeezing parameter as $r=1$, Fig. \ref{Fig.2} shows the
influence of the mean thermal photons numbers $n_{\mathrm{th,}1}$ and $n_{%
\mathrm{th,}2}$ on the dynamics of the Gaussian steerabilities $\mathcal{G}%
^{A\rightarrow B}$ and $\mathcal{G}^{B\rightarrow A}$, the steering
asymmetric $\mathcal{G}_{AB}^{\Delta }$ and entanglement $\mathcal{E}_{2}$.
The mean thermal occupations $n_{\mathrm{th,}1}$ and $n_{\mathrm{th,}2}$ are
fixed as : $n_{\mathrm{th,}1}=0.5$, $n_{\mathrm{th,}2}=1$ (panel (a)), $n_{%
\mathrm{th,}1}=1$, $n_{\mathrm{th,}2}=0.5$ (panel (b)), $n_{\mathrm{th,}1}=1$%
, $n_{\mathrm{th,}2}=1.2$ (panel (c)) and $n_{\mathrm{th,}1}=1$, $n_{\mathrm{%
th,}2}=1.5$ (panel (d)). As seen from Fig. \ref{Fig.2}, $\mathcal{G}%
^{A\rightarrow B}$, $\mathcal{G}^{B\rightarrow A}$ and $\mathcal{E}_{2}$
have the same time-evolution behavior. Indeed, the initial phase is a period
where $\mathcal{G}^{A\rightarrow B}$, $\mathcal{G}^{B\rightarrow A}$ and $%
\mathcal{E}_{2}$ are zero, exhibiting a time delay before a sudden birth,
which is analogous to the superradiance phenomenon. The second phase occurs
when the three measures follow a chronological hierarchy and gradual
build-up until a maximal value, and finally the third phase occurs when the
three measures start to diminish. Moreover, Fig. \ref{Fig.2} shows that the
influence of the asymmetric values of $n_{\mathrm{th,}1}$ and $n_{\mathrm{th,%
}2}$ is not only reflected on the time-generation of the steerabilities $%
\mathcal{G}^{A\rightarrow B}$ and $\mathcal{G}^{B\rightarrow A}$ but also on
their time-residence too. Fig. \ref{Fig.2} depicts also that steerable
states are always entangled as expected, but entangled states are not
necessarily steerable, which means that stronger quantum correlations are
required for achieving the steering than that for the entanglement. More
important, Fig. \ref{Fig.2} shows different situations where $\mathcal{G}%
^{A\rightarrow B}=0$, $\mathcal{G}^{B\rightarrow A}$ $>0$ and $\mathcal{E}%
_{2}>0$, which witnesses the existence of Gaussian one-way steering, i.e.,
the states of the two modes $A$ and $B$ are steerable only from $B $ to $A$
even though they are entangled. This behavior constitutes a genuine response
to the problem which has been discussed in \cite{Wiseman}. In addition, Fig. %
\ref{Fig.2} reveals that the two steerabilities $B\rightarrow A$ and $%
A\rightarrow B$ are strongly sensitive to the variations of $n_{\mathrm{th,}%
1}$ and $n_{\mathrm{th,}2}$ than entanglement, and they have a tendency to
disappear rapidly when the temperature increases.\newline
Now, fixing the mean thermal photons numbers as $n_{\mathrm{th,}1}=n_{%
\mathrm{th,}2}=1$, we discuss the dynamics of $\mathcal{G}^{A\rightarrow B}$%
, $\mathcal{G}^{B\rightarrow A}$ and $\mathcal{E}_{2}$ under influence of
the squeezing parameter $r$.
\begin{figure}[t]
\centerline{\includegraphics[width=0.5\columnwidth,height=4.5cm]{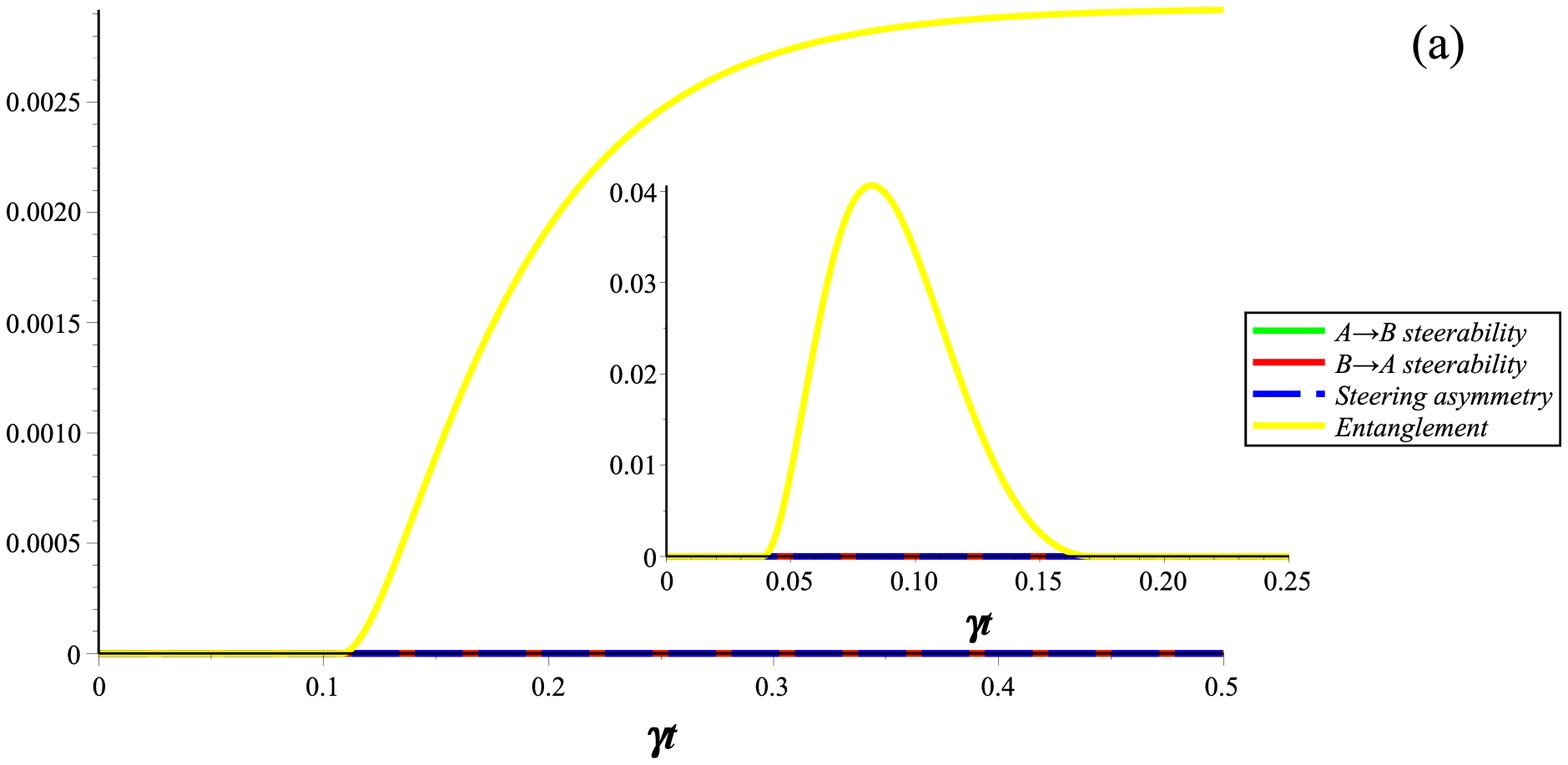}%
\includegraphics[width=0.5\columnwidth,height=4.5cm]{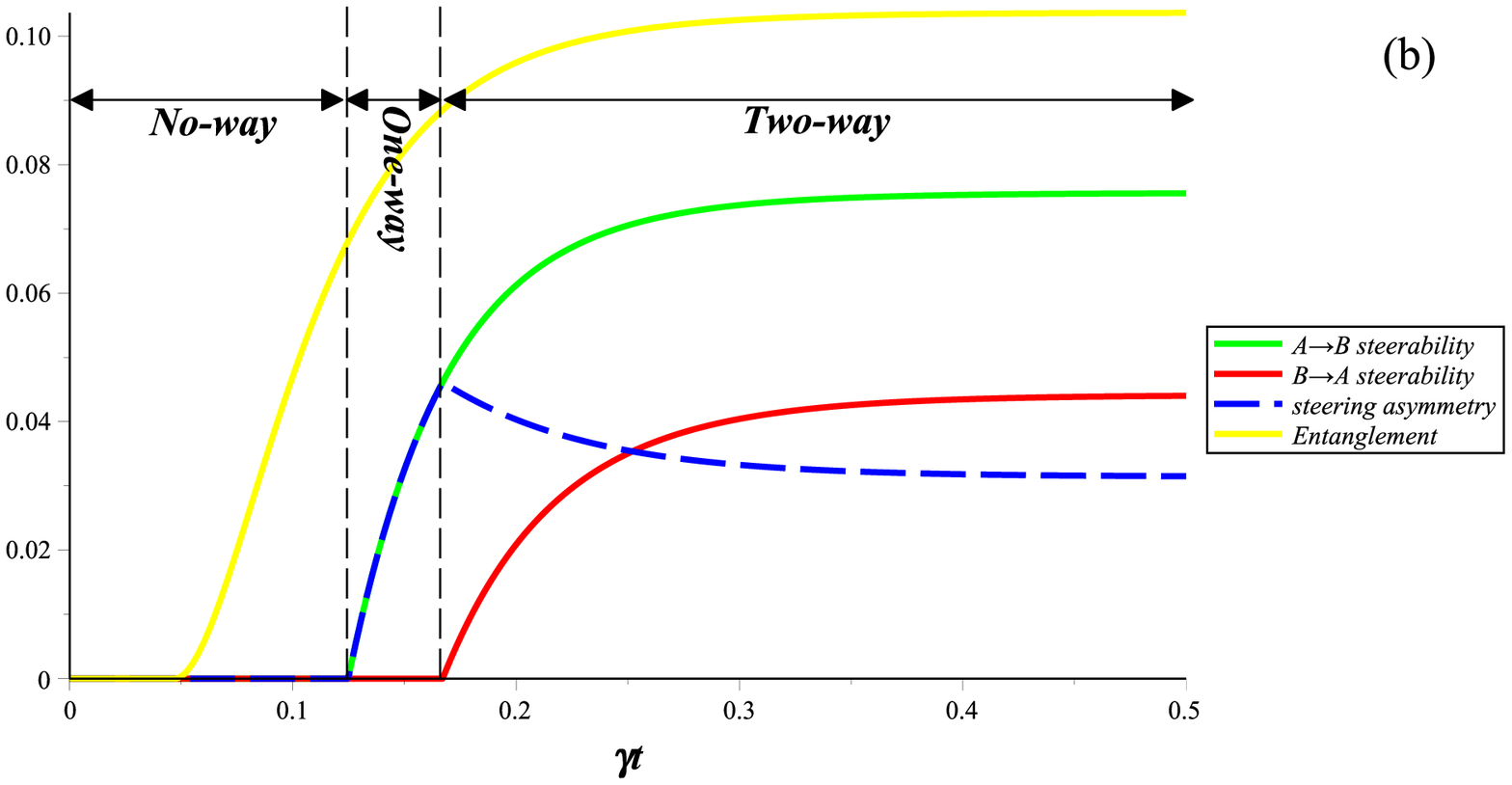}} %
\centerline{\includegraphics[width=0.5\columnwidth,height=4.5cm]{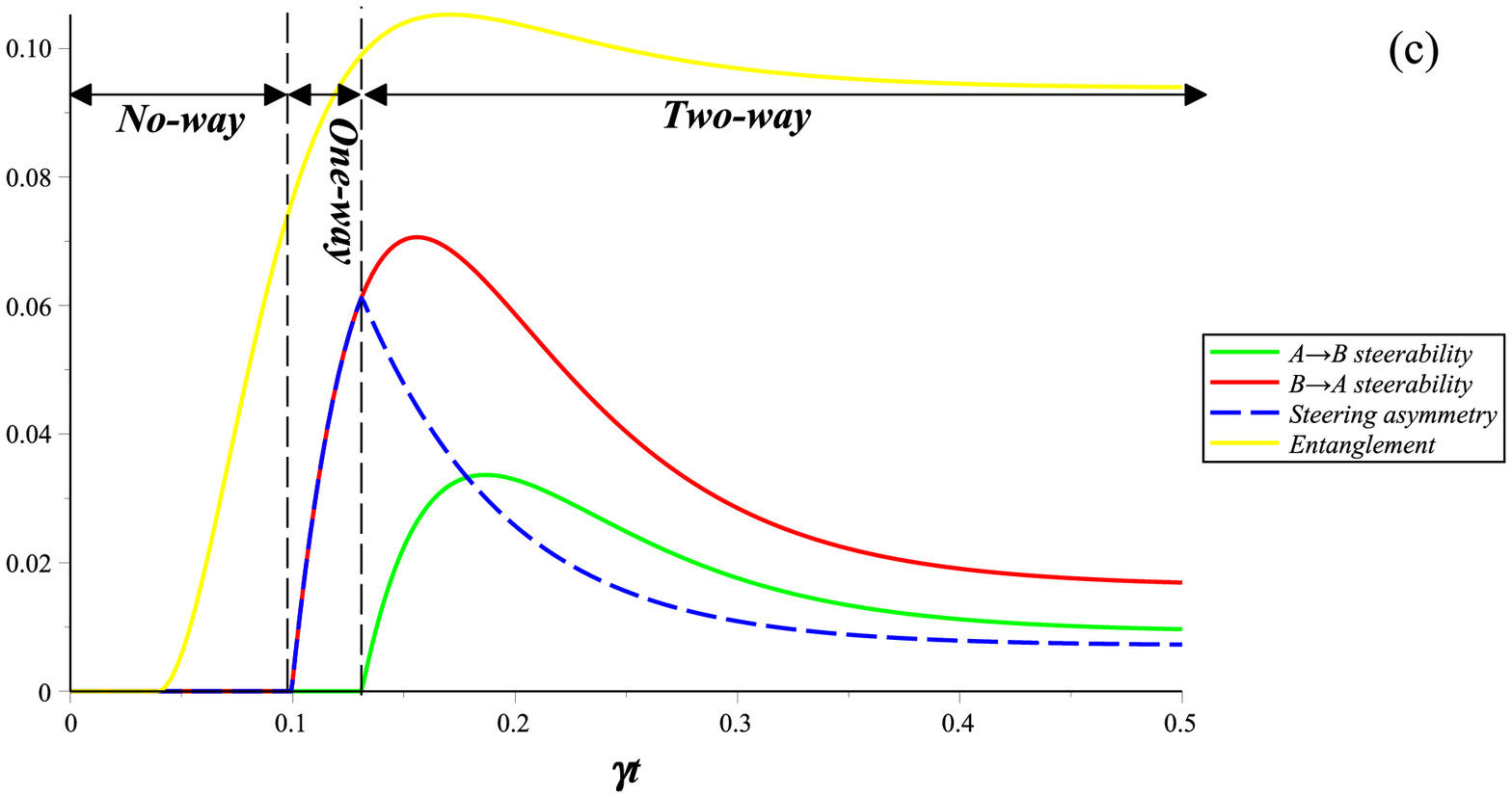}%
\includegraphics[width=0.5\columnwidth,height=4.5cm]{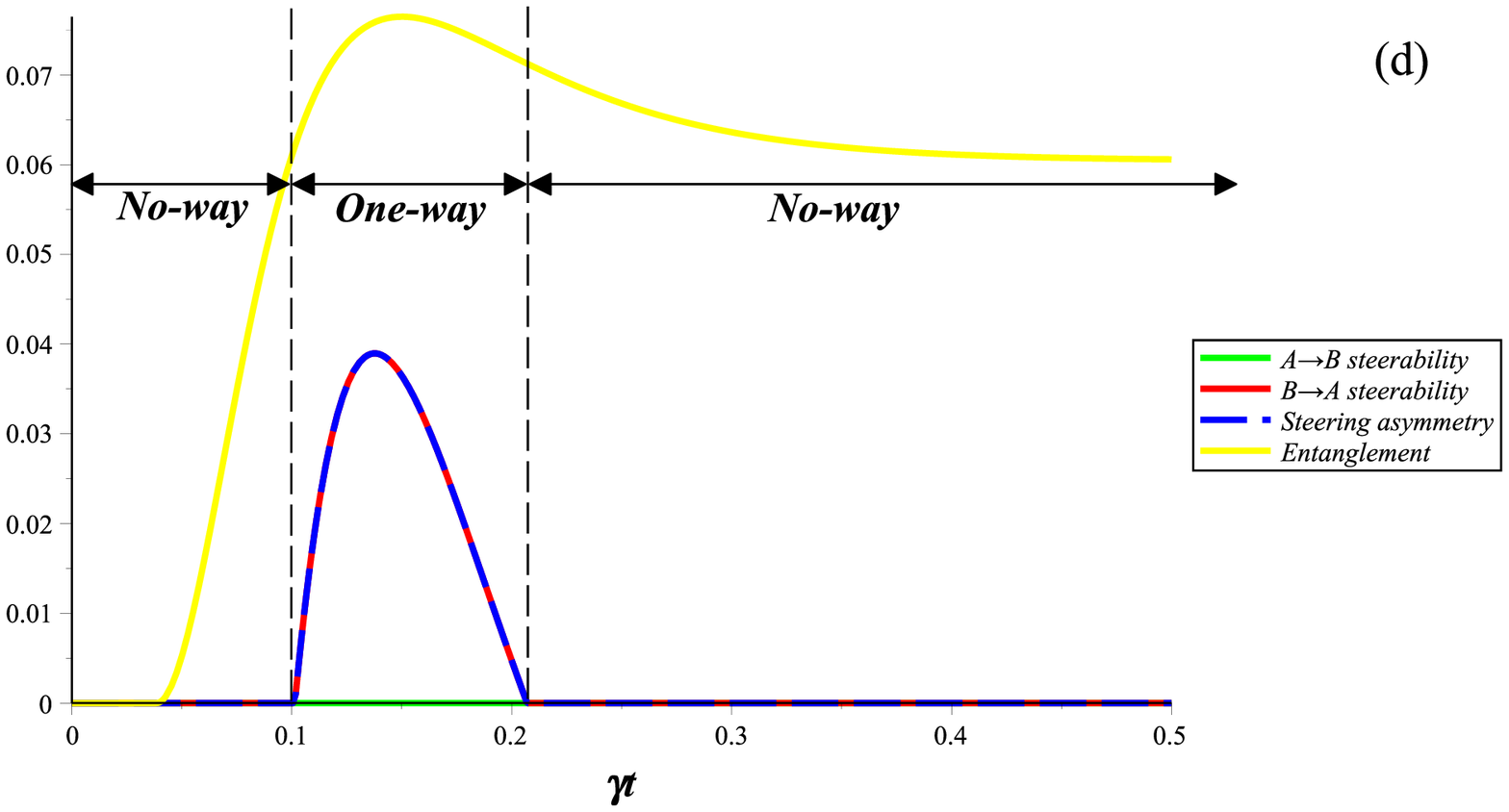}}
\caption{Plot of the Gaussian steering $\mathcal{G}^{A\rightarrow B}$ (green
solid line), $\mathcal{G}^{B\rightarrow A}$ (red solid line), the steering
asymmetry $\mathcal{G}_{AB}^{\Delta }$ (blue dashed line) and entanglement $%
\mathcal{E}_{2}$ (yellow solid line) of the two mechanical modes $A$ and $B$
as a function of the scaled time $\protect\gamma t$ for $\mathcal{C}_{1}=15$
and $\mathcal{C}_{2}=35$. We used $n_{\mathrm{th},1}=n_{\mathrm{th},2}=1$ as
values of the mean thermal photons numbers. The squeezing parameter $r$ is
fixed as : $r=0.1$ (panel (a)), $r=0.5$ (panel (b)), $r=1$ (panel (c)) and $%
r=1.1$ (panel (d)) and $r=1.7$ in the inset. Panel (d) shows a situation
where the states of the two mechanical modes are entangled (for $\protect%
\gamma t>0.05$); nevertheless they are straightforwardly steerable only in
one direction (from $B\rightarrow A$), which reflects genuinely the
asymmetry of quantum correlations between the modes $A$ and $B$. As shown
also in panel (a) and in the inset, entangled states are not necessarily
steerable, whereas steerable states are always entangled as depicted in the
panels (b), (c) and (d).}
\label{Fig.3}
\end{figure}
Firstly, like the results which have been presented in Fig. \ref{Fig.2}, we
see from Fig. \ref{Fig.3} that steerable states are always entangled,
whereas entangled states are not in general steerable. Moreover, Fig. \ref%
{Fig.3} shows that with gradual increase of the squeezing parameter $r$ : $%
r=0.1$ (panel (a)), $r=0.5$ (panel (b)), $r=1$ (panel (c)), $r=1.1$ (panel
(d)) and $r=1.7$ (in the inset), the squeezing has two opposite effects
(enhancement and degradation) on the behavior of $\mathcal{G}^{A\rightarrow
B}$, $\mathcal{G}^{B\rightarrow A}$ and $\mathcal{E}_{2}$. The enhancement
is due to the fact that the photon number in the two cavities increases
which enhances the optomechanical coupling by means of radiation pressure
and consequently leads to robust quantum correlations. However, in the
degradation period, the input thermal noise affecting each cavity becomes
important and more aggressive, causing the quantum correlation degradation.
This double-effect of the two-mode squeezed light can be understood based on
the fact that the reduced state of a two-mode squeezed light is a thermal
state having an average number of photons proportional to the squeezing
parameter $r$ \cite{Mauro}. On the other hand, comparing with entanglement,
it can be clearly seen from Fig. \ref{Fig.3} that quantum steering is
considerably sensitive to thermal noise induced by the gradual increasing of
$r$. Fig. \ref{Fig.3}(d) shows an interesting situation where the states of
the two mechanical modes $A$ and $B$ are entangled (for $\gamma t>0.05$);
nevertheless they are steerable only in one direction (from $B\rightarrow A$%
), which reflects genuinely the asymmetry of quantum correlations. Such a
property translates the fact that Alice and Bob can perform exactly the same
Gaussian measurements on their part of the entangled system, but obtain
different results. This can be explained by the asymmetry introduced in the
system and also by the definition of quantum steering in terms of the EPR
paradox \cite{Reid(2),Kogias}. Finally, all results depicted in Figs. \ref%
{Fig.2} and \ref{Fig.3} show that the Gaussian quantum steering is always
upper bounded by the Gaussian R\'{e}nyi-2 entanglement $\mathcal{E}_{2}$.
Moreover, the steering asymmetry $\mathcal{G}_{AB}^{\Delta }$ (see the blue
dashed-lines in Figs. \ref{Fig.2} and \ref{Fig.3}) is always less than $\ln
2 $, it is maximal when the state is nonsteerable in one way ($\mathcal{G}%
^{A\rightarrow B}>0$ and $\mathcal{G}^{B\rightarrow A}=0$ or $\mathcal{G}%
^{A\rightarrow B}=0$ and $\mathcal{G}^{B\rightarrow A}>0$) and it decreases
with increasing steerability in either way, which is consistent with the
literature \cite{Kogias}.

\section{Conclusions \label{sec4}}

Using the criterion proposed in \cite{Kogias}, dynamical Gaussian quantum
steering and its asymmetry of two mixed mechanical modes $A$ and $B$ have
been studied. A specific attention has been devoted to the dynamics of the
Gaussian one-way steerability. For this, a double-cavity optomechanical
system coupled to a common two-mode squeezed light has been employed. We
worked in the resolved sideband regime with high quality factor mechanical
oscillators. Eliminating adiabatically the optical cavities modes, we have
derived the explicit time-dependent expression of the covariance matrix (Eq.
(\ref{E24})) fully describing the mechanical fluctuations. In this way, we
have shown that it is possible to generate dynamical Gaussian quantum
steering via a quantum fluctuations transfer from the two-mode squeezed
light to the mechanical modes, whereas by an appropriate choice of the
environmental parameters (thermal occupations $n_{\mathrm{th},1}$, $n_{%
\mathrm{th},2}$ and squeezing $r$), Gaussian one-way steering can be
observed in different scenarios : (\textit{i}) Gaussian one-way steering has
been detected from $A\rightarrow B$ (see Fig. \ref{Fig.3}(b)) as well as
from $B\rightarrow A$ (see Fig. \ref{Fig.2} and Figs. \ref{Fig.3}(c)-\ref%
{Fig.3}(d)), (\textit{ii}) it has been observed from $B\rightarrow A$ during
two periods (see Figs. \ref{Fig.2}(c)-\ref{Fig.2}(d)), and finally (\textit{%
iii}) Gaussian one-way steering has occurred without two-way steering
behavior (see Fig. \ref{Fig.3}(d)). We have shown also that in some
circumstances which are governed by thermal effects, one can observe the
situation where the two mechanical modes are entangled, yet are
straightforwardly steerable only in one direction (see Fig. \ref{Fig.3}(d)),
which reflects genuinely the asymmetry of quantum correlations. On the other
hand, we have numerically compared the Gaussian steering of the two
mechanical modes $A$ and $B$ with their corresponding entanglement. Using
the Gaussian R\'{e}nyi-2 entropy as a measure of entanglement, we showed
that Gaussian steering is strongly sensitive to the thermal effects than
entanglement and always upper bounded by the Gaussian R\'{e}nyi-2
entanglement $\mathcal{E}_{2}$. Furthermore, we have found that the steering
asymmetry $\mathcal{G}_{AB}^{\Delta }$ is always less than $\ln 2$, it is
maximal when the state is nonsteerable in one way, and it decreases with
increasing steerability in either way, which is consistent with the
literature \cite{Kogias}. \newline
So, we believe that a Fabry-Perot double-cavity optomechanical system can be
of immediate practical interest in the investigation of Gaussian quantum
steering and its asymmetry between two mechanical modes. In addition, the
transfer of quantum fluctuations from two-mode squeezed light to mechanical
motions can be exploited to gain quantum advantages in implementing long
distance quantum protocols. We note also that an equivalent scheme can be
considered to study Gaussian steering between optical modes which may open a
new perspective in the context of quantum key distribution and in quantum
information science in general \cite{Branciard,Reid(3),Bancal}. Finally, it
will be interesting to investigate stationary Gaussian one-way steering in a
double-cavity optomechanical system using the criterion of Kogias \textit{et
al} \cite{Kogias}. We hope to report on this issue in a forthcoming work.
\section*{Acknowledgements}
The authors would like to thank David Vitali and Andrea Mari for many
useful discussions.
\section*{Author Contributions}
The authors contributed equally to this work.

\end{document}